\documentclass[11pt,a4paper]{article}

\usepackage{epsfig,amsmath,amssymb,cite}

\begin{document}

\title{Thin shells surrounding black holes in $F(R)$ gravity} 
\author{Ernesto F. Eiroa$^{1, 2}$\thanks{e-mail: eiroa@iafe.uba.ar}, Griselda Figueroa Aguirre$^{1}$\thanks{e-mail: gfigueroa@iafe.uba.ar}\\
{\small $^1$ Instituto de Astronom\'{\i}a y F\'{\i}sica del Espacio (IAFE, CONICET-UBA),}\\
{\small Casilla de Correo 67, Sucursal 28, 1428, Buenos Aires, Argentina}\\
{\small $^2$ Departamento de F\'{\i}sica, Facultad de Ciencias Exactas y 
Naturales,} \\ 
{\small Universidad de Buenos Aires, Ciudad Universitaria Pabell\'on I, 1428, 
Buenos Aires, Argentina}} 

\maketitle

\begin{abstract}

In this article, we consider spherical thin shells of matter surrounding black holes in $F(R)$ theories of gravity.  We study the stability of the static configurations under perturbations that conserve the symmetry. In particular, we analyze the case of charged shells  outside the horizon of non--charged black holes. We obtain that stable static thin shells are possible if the values of the parameters of the model are properly selected.\\

\noindent 
Keywords: Gravitation, Alternative gravity theories, Thin shells\\

\end{abstract}

\section{Introduction}\label{intro} 

The observations concerning the accelerated expansion of the Universe, the rotation curves of galaxies, and the anisotropy of the microwave background radiation can be explained within General Relativity by the presence of dark matter ($\sim$ 25 \%) and dark energy ($\sim$ 70 \%), besides the ordinary barionic matter ($\sim$ 5 \%). In the concordance or $\Lambda$CDM model, the dark energy contribution comes in the form of a cosmological constant $\Lambda$ and the cold dark matter in the form of non--relativistic fluid, supplemented by an inflationary scenario driven by a scalar field called the inflaton. Although successful, this model is not free of difficulties, such as the extremely small observed value of $\Lambda$ compared to the expected one if thought as originated from a vacuum energy in particle physics, or the unclear nature of dark matter (although several candidates exist). Other approaches can be adopted, such as modified gravity theories, in order to try to avoid these problems and explain the observed features of the Universe without dark matter and dark energy. Quantum gravity also provides motivation for modified gravity. One well known theory is $F(R)$ gravity \cite{dft,sofa,nojod} in which the Einstein-Hilbert Lagrangian is replaced by a function $F(R)$ of the Ricci scalar $R$. In recent years, several solutions of the field equations in $F(R)$ gravity have been found, including static and spherically symmetric black holes \cite{sofa,bhfr1,bhfr2,bhfr3,moon,bhfr4}, traversable wormholes \cite{whfr}, and branes \cite{branefr}.

The Darmois--Israel junction conditions \cite{daris} provide the tools for matching two solutions across a hypersurface in General Relativity. These conditions allow for the study of thin shells of matter, by relating the energy--momentum tensor at the joining hypersurface with the space--times at both sides of it. The formalism has been broadly adopted in many different scenarios due to its flexibility and simplicity; among them it is used to model vacuum bubbles and thin layers around black holes \cite{sh1,sh2,sh3}, gravastars \cite{gravstar1,gravstar2}, and wormholes \cite{wh1,wh2,wh3,whcil}. In the case of highly symmetric configurations, the stability analysis is usually easy to perform, at least for perturbations preserving the symmetry. 

The junction conditions in  $F(R)$ theories \cite{dss,js1} are more restrictive than in General Relativity. For non--linear $F(R)$, the trace of the second fundamental form should always be continuous at the matching hypersurface \cite{js1}. Except in the quadratic case, the curvature scalar $R$ should also be continuous there \cite{js1}. In quadratic $F(R)$ gravity, the hypersurface has in general, in addition to the standard energy--momentum tensor, an external energy flux vector, an external scalar pressure (or tension), and another energy--momentum contribution resembling classical dipole distributions. All these contributions have to be present \cite{js1,js2-3} in order to have a divergence--free energy--momentum tensor, which guarantees local conservation. It was recently shown that these features are shared by any theory with a quadratic lagrangian \cite{js4}. The junction conditions in $F(R)$ were recently applied to the construction of thin-shell wormholes \cite{eirfig,whfrts} and bubbles \cite{bb1,gfa18}. A particularly interesting example of a pure double layer in quadratic $F(R)$ was found \cite{bb1}.

In this work, we construct spherical thin shells surrounding  non-charged black holes by using the junction conditions in $F(R)$ gravity and we analyze the stability of the static configurations under perturbations that preserve the symmetry. In Sec. 2, we review the general formalism for spherical geometries with constant curvature scalars at both sides of the shell. In Sec. 3, we perform the construction and we study the stability of charged thin shells outside the black hole event horizon. Finally, in Sec. 4, we present the conclusions of the paper. We adopt a system of units in which $c=G=1$, with $c$ the speed of light and $G$ the gravitational constant.

\section{Spherical thin shells: construction and stability}

We begin by reviewing the formalism for spherical thin shells in four dimensional $F(R)$ gravity  introduced in Ref. \cite{gfa18}. We consider a manifold composed of two regions with a constant curvature scalar in each, separated by a thin shell of matter. For this purpose, we take two different spherically symmetric solutions in $F(R)$ gravity, with metrics
\begin{equation} 
ds^2=-A_{1,2} (r_{1,2}) dt_{1,2}^2+A_{1,2} (r_{1,2})^{-1} dr_{1,2}^2+r_{1,2}^2(d\theta^2 + \sin^2\theta d\varphi^2),
\label{metric}
\end{equation}
where $r_{1,2}>0$ are the radial coordinates corresponding to each geometry, and $0\le \theta \le \pi$ and $0\le \varphi<2\pi $ are the angular coordinates. We proceed with the construction of a new manifold by selecting a radius $a \equiv a_1=a_2$ and cutting two regions $\mathcal{M}_1$ and $\mathcal{M}_2$ defined as the inner $0\le r_1 \leq a$ and the outer $r_2\geq a$ parts of the geometries 1 and 2, respectively. These regions are pasted to one another at the surface $\Sigma $ with radius $a$. This construction results in the spacetime $\mathcal{M}=\mathcal{M}_1 \cup \mathcal{M}_2$, with the inner zone corresponding to $\mathcal{M}_1$ and the exterior one to $\mathcal{M}_2$. The angular coordinates have been naturally identified everywhere from the beginning. We define a new global radial coordinate $r\in [0,+\infty)$ by identifying $r$ with $r_1$ in $\mathcal{M}_1$ and with  $r_2$ in $\mathcal{M}_2$, respectively. Then, the global coordinates are $X^{\alpha }_{1,2} = (t_{1,2},r,\theta, \varphi)$, while on the surface $\Sigma $, corresponding to $G(r)\equiv r-a =0$, we adopt the coordinates $\xi ^{i}=(\tau ,\theta,\varphi )$, with $\tau $ the proper time. In what follows, we take the radius of the surface $\Sigma$  as a function $a(\tau)$ of the proper time. The equality of the proper time at the sides of the shell requires that
\begin{equation*}
\frac{dt_{1,2}}{d\tau} = \frac{\sqrt{A_{1,2}(a) + \dot{a} ^2}}{A_{1,2}(a)} ,
\end{equation*}
in which the free signs were fixed by choosing all times $t_{1,2}$ and $\tau$ to run to the future, and $\dot{a}$ is the proper time derivative of $a$. We denote the first fundamental form by $h_{\mu \nu}$, the second fundamental form (or extrinsic curvature) by $K_{\mu \nu}$, and the unit normals at the surface $\Sigma$ by $n^{1,2}_\gamma$ (pointing from $\mathcal{M}_1$ to $\mathcal{M}_2$).  The first fundamental form associated with the two sides of the shell is given by
\begin{equation}
h^{1,2}_{ij}= \left. g^{1,2}_{\mu\nu}\frac{\partial X^{\mu}_{1,2}}{\partial\xi^{i}}\frac{\partial X^{\nu}_{1,2}}{\partial\xi^{j}}\right| _{\Sigma },
\end{equation}
and the second fundamental form has components
\begin{equation}
K_{ij}^{1,2 }=-n_{\gamma }^{1,2 }\left. \left( \frac{\partial ^{2}X^{\gamma
}_{1,2} } {\partial \xi ^{i}\partial \xi ^{j}}+\Gamma _{\alpha \beta }^{\gamma }
\frac{ \partial X^{\alpha }_{1,2}}{\partial \xi ^{i}}\frac{\partial X^{\beta }_{1,2}}{
\partial \xi ^{j}}\right) \right| _{\Sigma },
\label{sff}
\end{equation}
with the unit normals ($n^{\gamma }n_{\gamma }=1$) determined by
\begin{equation}
n_{\gamma }^{1,2 }=\left\{ \left. \left| g^{\alpha \beta }_{1,2}\frac{\partial G}{\partial
X^{\alpha }_{1,2}}\frac{\partial G}{\partial X^{\beta }_{1,2}}\right| ^{-1/2}
\frac{\partial G}{\partial X^{\gamma }_{1,2}} \right\} \right| _{\Sigma }.
\end{equation}
We adopt at the surface $\Sigma $ the orthonormal basis $\{ e_{\hat{\tau}}=e_{\tau }, e_{\hat{\theta}}=a^{-1}e_{\theta }, e_{\hat{\varphi}}=(a\sin \theta )^{-1} e_{\varphi }\} $. Then, for the geometry given by (\ref{metric}), the first fundamental form results $h^{1,2}_{\hat{\imath}\hat{\jmath}}= \mathrm{diag}(-1,1,1)$, the unit normals read
\begin{equation} 
n_{\gamma }^{1,2}= \left(-\dot{a},\frac{\sqrt{A_{1,2}(a)+\dot{a}^2}}{A_{1,2}(a)},0,0 \right),
\end{equation}
and the second fundamental form non--null components are
\begin{equation} 
K_{\hat{\theta}\hat{\theta}}^{1,2}=K_{\hat{\varphi}\hat{\varphi}}^{1,2
}=\frac{1}{a}\sqrt{A_{1,2} (a) +\dot{a}^2}
\label{e4}
\end{equation}
and
\begin{equation} 
K_{\hat{\tau}\hat{\tau}}^{1,2 }=-\frac{A '_{1,2}(a)+2\ddot{a}}{2\sqrt{A_{1,2}(a)+\dot{a}^2}},
\label{e5}
\end{equation}
where the prime on $A_{1,2}(r)$ represents the derivative with respect to $r$. 

From now on, we denote with a prime on $F(R)$ the derivative with respect to the curvature scalar $R$ and the jump  of any quantity $\Upsilon $ across $\Sigma$ by $[\Upsilon ]\equiv (\Upsilon ^{2}-\Upsilon  ^{1})|_\Sigma $. The junction formalism in $F(R)$ gravity theories provides the conditions that should be fulfilled at $\Sigma $. One of them is the continuity of the first fundamental form i.e.  $[h_{\mu \nu}]=0$. It is straightforward to verify that this condition is satisfied by our construction. Another one is the continuity of the trace of the second fundamental form, i.e. $[K^{\mu}_{\;\; \mu}]=0$, which by using Eqs. (\ref{e4}) and (\ref{e5}), takes the form
\begin{equation} 
-\frac{2a\ddot{a}+a A_{1}'(a) + 4(A_{1}(a)+\dot{a}^2)}{\sqrt{A_{1}(a)+\dot{a}^2}}+\frac{2a\ddot{a}+a A_{2}'(a) + 4(A_{2}(a)+\dot{a}^2)}{\sqrt{A_{2}(a)+\dot{a}^2}}=0.
\label{CondGen}
\end{equation}
When $F'''(R) \neq 0$ the continuity of $R$ across the $\Sigma$ is also required i.e.  $[R]=0$. The field equations at $\Sigma$ read \cite{js1}
\begin{equation} 
\kappa S_{\mu \nu}=-F'(R)[K_{\mu \nu}]+ F''(R)[\eta^\gamma \nabla_\gamma R]  h_{\mu \nu}, \;\;\;\; n^{\mu}S_{\mu\nu}=0,
\label{LanczosGen}
\end{equation}
with $\kappa =8\pi $ and $S_{\mu \nu}$ the energy--momentum tensor at the shell. If $F'''(R) = 0$ (quadratic case), the curvature scalar can be discontinuous at $\Sigma $, and the field equations adopt the form \cite{js1}
\begin{equation}
\kappa S_{\mu \nu} =-[K_{\mu\nu}]+2\alpha \left( [n^{\gamma }\nabla_{\gamma}R] h_{\mu\nu}-[RK_{\mu\nu}] \right), \;\;\;\; n^{\mu}S_{\mu\nu}=0;
\label{LanczosQuad}
\end{equation}
supplemented by three other contributions: an external energy flux vector
\begin{equation}
\kappa\mathcal{T}_\mu=-2\alpha \bar{\nabla}_\mu[R],  \qquad  n^{\mu}\mathcal{T}_\mu=0,
\label{Tmu}
\end{equation}
with $\bar{\nabla }$ the intrinsic covariant derivative on $\Sigma$, an external scalar pressure or tension
\begin{equation}
\kappa\mathcal{T}=2\alpha [R] K^\gamma{}_\gamma ,
\label{Tg}
\end{equation}
and a two-covariant symmetric tensor distribution 
\begin{equation}
\kappa \mathcal{T}_{\mu \nu}=\nabla_{\gamma } \left( 2\alpha [R] h_{\mu \nu } n^{\gamma } \delta ^{\Sigma }\right),
\label{dlay1}
\end{equation}
with $\delta ^{\Sigma }$ the Dirac delta on $\Sigma $. This last expression admits an equivalent form 
\begin{equation}
\kappa \left<\mathcal{T}_{\mu \nu},\Psi ^{\mu \nu } \right> = -\int_\Sigma 2\alpha[R] h_{\mu \nu }  n^\gamma\nabla_\gamma \Psi ^{\mu \nu },
\label{dlay2}
\end{equation}
for any test tensor field $\Psi ^{\mu\nu}$. 
In quadratic $F(R)$, the shell has, in addition to the standard energy--momentum tensor $S_{\mu \nu}$, an external energy flux vector $\mathcal{T} _{\mu}$, an external scalar tension/pressure $\mathcal{T} $, and a double layer tensor distribution $\mathcal{T}_{\mu \nu}$ of Dirac ``delta prime'' type, having a resemblance with classical dipole distributions. All these contributions are necessary in order to ensure the energy--momentum tensor to be divergence--free, a condition that is required for conservation locally  \cite{js1}.

\subsection{The same constant curvature scalar $R_0$}\label{tswh}

We firstly study the case with a constant curvature scalar $R_0$ at both sides of $\Sigma$. Then, the condition $[R]=0$ is automatically fulfilled when required, and Eqs. (\ref{LanczosGen}) and (\ref{LanczosQuad}) both simplify to give
\begin{equation} 
\kappa S_{\mu \nu}=-F'(R_0)[K_{\mu \nu}].
\label{fieldeq}
\end{equation}
In quadratic $F(R)$ the contributions $\mathcal{T} $, $\mathcal{T} _{\mu}$ and $\mathcal{T}_{\mu \nu}$ are all zero due to their proportionality to $[R]$. The energy--momentum tensor in the orthonormal basis takes the form $S_{_{\hat{\imath}\hat{\jmath} }}={\rm diag}(\sigma ,p_{\hat{\theta}},p_{\hat{\varphi}})$, with $\sigma$ the surface energy density and $p_{\hat{\theta}}=p_{\hat{\varphi}}=p$ the transverse pressures, so from Eq. (\ref{fieldeq}) we find that
\begin{equation} 
\sigma= \frac{F'(R_0)}{2\kappa}\left( \frac{2\ddot{a}+A_{2}'(a)}{\sqrt{A_{2}(a)+\dot{a}^2}}-\frac{2\ddot{a}+A_{1}'(a)}{\sqrt{A_{1}(a)+\dot{a}^2}}\right)
\label{e9}
\end{equation}
and
\begin{equation}
p=\frac{- F'(R_0)}{\kappa a}\left( \sqrt{A_{2}(a)+\dot{a}^2}-\sqrt{A_{1}(a)+\dot{a}^2}\right).
\label{e10}
\end{equation}
In $F(R)$ gravity, the inequality $F'(R)>0$ implies that the effective Newton constant $G_{eff} = G/F'(R) = 1/F'(R)$ is positive \cite{bhfr2}, so preventing, from a quantum point of view, the graviton to be a ghost. An interesting discussion about this issue, within a wormhole scenario, is presented in Ref. \cite{bronnikov}. We assume the absence of ghosts, so we demand that $F'(R_0)>0$ from now on. Normal matter satisfies the weak energy condition, which in the orthonormal frame requires that $\sigma \geq 0$ and $\sigma + p \geq 0$; if it does not, the matter is exotic. From Eqs. (\ref{CondGen}), (\ref{e9}), and (\ref{e10}) we obtain the equation of state
\begin{equation}
\sigma -2p=0.
\label{e11} 
\end{equation} 
By taking the time derivative of the equation of state and using Eqs. (\ref{e9}) and (\ref{e10}), we obtain the conservation equation
\begin{equation}
\frac{d(\sigma \mathcal{A})}{d\tau}+p\frac{d \mathcal{A}}{d\tau}=0,
\label{conservacion2}
\end{equation}
where $\mathcal{A}=4\pi a^2$ is the area of the shell. The first term in this equation represents the internal energy change while the second one is the work done by the internal forces at the shell.

For static configurations with a constant radius $a_0$, Eq. (\ref{CondGen}) reduces to
\begin{equation} 
-\frac{a_0 A_{1}'(a_0) + 4A_{1}(a_0)}{\sqrt{A_{1}(a_0)}}+\frac{a_0 A_{2}'(a_0)+ 4A_{2}(a_0)}{\sqrt{A_{2}(a_0)}}=0.
\label{CondEstatico}
\end{equation}
The surface energy density $\sigma _0$ and the pressure $p _0$ take the form
\begin{equation} 
\sigma_0=\frac{F'(R_0)}{2\kappa}\left( \frac{A_{2}'(a_0)}{\sqrt{A_{2}(a_0)}}-\frac{A_{1}'(a_0)}{\sqrt{A_{1}(a_0)}}\right)
\label{e13}
\end{equation}
and
\begin{equation}
p_0= \frac{-F'(R_0)}{\kappa a_0}\left( \sqrt{A_{2}(a_0)}-\sqrt{A_{1}(a_0)}\right),
\label{e14}
\end{equation}
respectively; they fulfill the equation of state $\sigma_0 - 2 p_0=0$.

We analyze the stability of static solutions under perturbations preserving the spherical symmetry. By using that $\ddot{a}= (1/2)d(\dot{a}^2)/da$ and with the definition $z=\sqrt{A_{2}(a)+\dot{a}^2}-\sqrt{A_{1}(a)+\dot{a}^2}$, we can rewrite Eq. (\ref{CondGen}) to give $az'(a)+2z(a)=0$; so by solving this equation we find an expression for $\dot{a}^{2}$ in terms of an effective potential
\begin{equation}
\dot{a}^{2}=-V(a),
\label{condicionPot}
\end{equation}
where 
\begin{equation}
V(a)= -\frac{a_{0}^4\left(\sqrt{A_{2}(a_{0})}-\sqrt{A_{1}(a_{0})}\right)^2}{4a^4} +\frac{A_{1}(a)+A_{2}(a)}{2} -\frac{a^4 \left(A_{2}(a)-A_{1}(a)\right)^{2}}{4 a_{0}^4\left(\sqrt{A_{2}(a_{0})}-\sqrt{A_{1}(a_{0})}\right)^2}.
\label{potencial}
\end{equation}
It is easy to see that $V(a_0)=0$ and by using Eq. (\ref{CondEstatico}) that $V'(a_0)=0$. The second derivative of the potential at $a_0$ reads
\begin{eqnarray}
V''(a_0)&=& -\frac{5 \left(\sqrt{A_{2}(a_{0})}-\sqrt{A_{1}(a_{0}})\right)^{2}}{a_{0}^2}  -\frac{3\left(\sqrt{A_{1}(a_{0})}+\sqrt{A_{2}(a_{0}})\right)^{2}}{a_{0}^2}\nonumber \\
&&-\frac{\left(A_{2}'(a_{0})-A_{1}'(a_{0})\right)^2}{2\left(\sqrt{A_{2}(a_{0})}-\sqrt{A_{1}(a_{0})}\right)^{2}}
-\frac{4\left(\sqrt{A_{1}(a_{0})}+\sqrt{A_{2}(a_{0}})\right)^{2}\left(A_{2}'(a_{0})-A_{1}'(a_{0})\right)}{a_{0}\left(A_2(a_0)-A_1(a_0)\right)}  \nonumber \\
&&+\frac{A_{1}''(a_{0})+A_{2}''(a_{0})}{2}
-\frac{\left(\sqrt{A_{1}(a_{0})}+\sqrt{A_{2}(a_{0})}\right)^{2}\left(A_{2}''(a_{0})-A_{1}''(a_{0})\right)}{2\left(A_{2}(a_{0})-A_{1}(a_{0})\right)}.
\label{potencial2der}
\end{eqnarray}
A static configuration having a radius $a_0$ is stable if and only if $V''(a_0)>0$, corresponding to a minimum of the potential. 

\subsection{Different and constant curvature scalars $R_1$ and $R_2$}

Now we take two different curvature scalars $R_1$ and $R_2$ at the sides of the shell $\Sigma $. The jump $[R]\neq 0$ restricts our study to the quadratic case $F(R)=R-2\Lambda+\alpha R^{2}$ case. Consequently, we only demand the continuity of the first fundamental form and of the trace of the second fundamental form, i.e. $[h_{\mu \nu}]=0$ and $[K^{\mu}_{\;\; \mu}]=0$.  
We proceed as above, but now with constant $R_1 \neq R_2$. The shell radius $a$ has to satisfy Eq. (\ref{CondGen}). From Eq. (\ref{LanczosQuad}), the field equations take the form
\begin{equation}
\kappa S_{\mu \nu} =-[K_{\mu\nu}]-2\alpha[RK_{\mu\nu}];
\label{Quad_LanczosGen2}
\end{equation}
then, using that $S_{\hat{\imath}\hat{\jmath} }={\rm diag}(\sigma ,p,p)$ in the orthonormal basis, we find that the energy density and the transverse pressure at the shell are
\begin{equation} 
\sigma= -\frac{2 \ddot{a}+A'_{1}(a)}{2\kappa\sqrt{A_{1}(a)+\dot{a}^2}}\left(1+2\alpha R_{1}\right)+\frac{2 \ddot{a}+A'_{2}(a)}{2\kappa\sqrt{A_{2}(a)+\dot{a}^2}}\left(1+2\alpha R_{2}\right) ,
\label{enden}
\end{equation}
\begin{equation}
p=\frac{\sqrt{A_{1}(a)+\dot{a}^2}}{\kappa a} \left(1+2\alpha R_{1} \right)-\frac{\sqrt{A_{2}(a)+\dot{a}^2}}{\kappa a} \left(1+2\alpha R_{2} \right),
\label{pres}
\end{equation}
respectively. As it was explained before, we assume that $F'(R_1)=1+ 2\alpha R_1>0$ and $F'(R_2)=1+ 2\alpha R_2>0$, in order to avoid the presence of ghosts. The matter is normal at $\Sigma$ if it satisfies the weak energy condition. From Eq. (\ref{Tmu}) we obtain that $\mathcal{T}_\mu =0$ and from to Eq. (\ref{Tg}) the external scalar tension/pressure $\mathcal{T}$ is given by
\begin{equation}
\mathcal{T}=\frac{2\alpha [R]}{\kappa \sqrt{A_{1}(a)+\dot{a}^2}}\left(\ddot{a}+\frac{A'_{1}(a)}{2}+\frac{2}{a}\left(A_{1}(a)+\dot{a}^2\right)\right),
\label{T}
\end{equation}
or, using Eq. (\ref{CondGen}), by the expression
\begin{equation} 
\mathcal{T}=-\frac{ 2a\ddot{a}+a A_{1}'(a) + 4 \left(A_{1}(a) +\dot{a}^2\right)}{\kappa a\sqrt{A_{1}(a) +\dot{a}^2}}\alpha R_{1}+\frac{2a\ddot{a}+a A_{2}'(a) + 4 \left(A_{2}(a) +\dot{a}^2\right)}{\kappa a\sqrt{A_{2}(a) +\dot{a}^2}}\alpha R_{2}.
\label{Trewrit}
\end{equation}
From Eqs. (\ref{enden}), (\ref{pres}), and (\ref{Trewrit}) we obtain the equation of state relating $\sigma$, $p$, and $\mathcal{T}$ 
\begin{equation}
\sigma-2p=\mathcal{T}.
\label{state}
\end{equation}
By taking the time derivative of Eq. (\ref{state}) and with the help of Eqs. (\ref{enden}) and (\ref{pres}), we readily find the generalized conservation equation
\begin{equation} 
\frac{d}{d\tau}(\mathcal{A}\sigma )+ p \frac{d\mathcal{A}}{d\tau} = \mathcal{A} \frac{d\mathcal{T}}{d\tau},
\label{continuity-bis}
\end{equation}
with the area $\mathcal{A}$ defined above. In the left hand side of this equation, the first term is thought as the change in the total energy of the shell, the second one as the work done by the internal pressure, while the right hand side represents an external flux. The double layer distribution $\mathcal{T}_{\mu \nu }$, should satisfy Eq. (\ref{dlay2}), which in our case adopts the form
\begin{equation} 
\langle \mathcal{T}_{\mu \nu } ,\Psi ^{\mu \nu } \rangle= - \int_{\Sigma}  \mathcal{P}_{\mu \nu } \left( n^t \nabla _t \Psi ^{\mu \nu} + n^r \nabla _r \Psi ^{\mu \nu } \right),
\label{dlay3}
\end{equation}
for any test tensor field $\Psi ^{\mu \nu }$. The components in the orthonormal basis of the double layer distribution strength are
\begin{equation}
- \mathcal{P}_{\tau \tau} =\mathcal{P}_{\hat{\theta}\hat{\theta} }=\mathcal{P}_{\hat{\varphi}\hat{\varphi} }=\frac{2\alpha[R]}{\kappa },
\label{dlsob}
\end{equation}
which only depend on $\alpha $ and $[R]$; as a consequence, the dependence of $\mathcal{T}_{\hat{\imath}\hat{\jmath} }$ with the metric comes from the unit normal and the covariant derivative.  

In the static configurations, the radius $a_0$ should fulfill Eq. (\ref{CondEstatico}), and from Eqs. (\ref{enden}),  (\ref{pres}), and  (\ref{Trewrit}), the surface energy density $\sigma _0$, the pressure $p_0$, and the external tension/pressure $\mathcal{T}_0$ read
\begin{equation} 
\sigma_0=- \frac{A'_{1}(a_{0})}{2\kappa\sqrt{A_{1}(a_{0})}}\left(1+2\alpha R_{1}\right)+\frac{A'_{2}(a_{0})}{2\kappa\sqrt{A_{2}(a_{0})}}\left(1+2\alpha R_{2}\right) ,
\label{enden0}
\end{equation}
\begin{equation}
p_0=\frac{\sqrt{A_{1}(a_{0})}}{\kappa a_{0}} \left(1+2\alpha R_{1} \right)-\frac{\sqrt{A_{2}(a_{0})}}{\kappa a_{0}} \left(1+2\alpha R_{2} \right),
\label{pres0}
\end{equation}
and
\begin{equation}
\mathcal{T}_0=-\frac{ a_0 A_{1}'(a_0) + 4 A_{1}(a_0)}{\kappa a_0 \sqrt{A_{1}(a_0)}}\alpha R_{1} +\frac{a_0 A_{2}'(a_0)+4 A_{2}(a_0)}{\kappa a_0 \sqrt{A_{2}(a_0)}}\alpha R_{2},
\label{T0}
\end{equation}
respectively. The equation of state becomes $\sigma_0 -2 p_0 =  \mathcal{T}_0$. The shell has a null external energy flux vector $\mathcal{T} _{\mu}^{(0)}$ and a non--null double layer distribution $\mathcal{T}_{\mu \nu }^{(0)}$ given by Eq. (\ref{dlay3}), with $n^t \nabla _t \Psi ^{\mu \nu} =0$ and the strength shown in Eq. (\ref{dlsob}). 

Following the same procedure of the previous sub--section, the stability of the static configurations is found from Eq. (\ref{condicionPot}) in terms of the potential of Eq. (\ref{potencial}), with its second derivative given by Eq. (\ref{potencial2der}); again $V''(a_0)>0$ correspond to the stable ones.

\section{Charged thin shells}\label{charge}

We start from the action corresponding to $F(R)$ gravity coupled to Maxwell electrodynamics
\begin{equation}
S=\frac{1}{2 \kappa}\int d^4x \sqrt{|g|} (F(R)-\mathcal{F}_{\mu\nu}\mathcal{F}^{\mu\nu}),
\label{action} 
\end{equation} 
where $g=\det (g_{\mu \nu})$ is the determinant of the metric tensor and $\mathcal{F}_{\mu \nu }=\partial _{\mu }\mathcal{A}_{\nu } -\partial _{\nu }\mathcal{A}_{\mu }$ is the electromagnetic tensor. In the metric formalism, the field equations obtained from this action, considering an electromagnetic potential $\mathcal{A}_{\mu}=(\mathcal{V}(r),0,0,0)$ and a constant curvature scalar $R$, admit a spherically symmetric solution  given by Eq. (\ref{metric}), in which the metric function \cite{bhfr2,moon} has the form 
\begin{equation} 
A (r) = 1-\frac{2M}{r}+\frac{Q^2}{ F'(R) r^2}-\frac{R r^2}{12},
\label{A_metric}
\end{equation}
with $Q$ the charge and $M$ the mass. The electromagnetic potential is $\mathcal{V}(r)=-Q/r$ and the cosmological constant is related with the curvature scalar by $4\Lambda = R$. 

\subsection{Curvature scalar $R_0$ at both sides}

For the construction of the thin shell $\Sigma $, we take the metric function given by Eq. (\ref{A_metric}), with mass $ M_1 \neq 0 $ and null charge for the internal region $\mathcal{M}_1 $, and mass $ M_2 \neq 0 $ and charge $ Q $ for the external region $ \mathcal{M}_2 $. At both sides of $\Sigma $ we adopt the same value $ R_0 $ for the curvature scalar. Therefore, the metric functions read
\begin{equation} 
A _1(r) = 1-\frac{2M_1}{r}-\frac{R_0 r^2}{12},
\label{A1_metric}
\end{equation}
for the internal zone, and
\begin{equation} 
A _2(r) = 1-\frac{2M_2}{r}+\frac{Q^2}{ F'(R_0) r^2}-\frac{R_0 r^2}{12}.
\label{A2_metric}
\end{equation}
for the external one. The possible horizons result from the zeros of the expressions $ A_{1} (r) $ and $ A_{2} (r) $. Both metrics are singular at $ r = 0 $. When $ A_{1} (r) = 0 $ we get a polynomial of third degree if $ R_0 \neq 0 $, its real and positive roots correspond to the radii of the different horizons. For $ R_0 \leqslant 0 $ there is only the presence of an event horizon. When $ 0 < R_0 < 4 / (9M_1 ^ 2) $ there is a cosmological horizon in addition to the event horizon. For the metric function $ A_{2} (r)$, the horizons are determined by the solutions of the quadratic equation when $ R_0 = 0 $, while for $ R_0 \neq 0 $ they are given by the real and positive roots of a  fourth degree polynomial. The critical charge value $ Q_c $ has an important role in the study of the solutions, since it determines the number of horizons of the geometry. If $ R_0 <0 $ and $0 < | Q | <Q_c $ there are two horizons, the internal and the event ones. When $ | Q | = Q_c $, they fuse into one, and if $ | Q |> Q_c $ only a naked singularity is left. For $ R_0 > 0 $ the metric has a cosmological horizon. Besides it, when $ 0 < | Q | < Q_c $  there exist an internal and an event horizons, if $ | Q | = Q_c $ both merged into one, and they finally disappear when $ | Q |> Q_c $, so that there is a naked singularity at the origin. 

\begin{figure}[t!]
\centering
\includegraphics[width=0.8\textwidth]{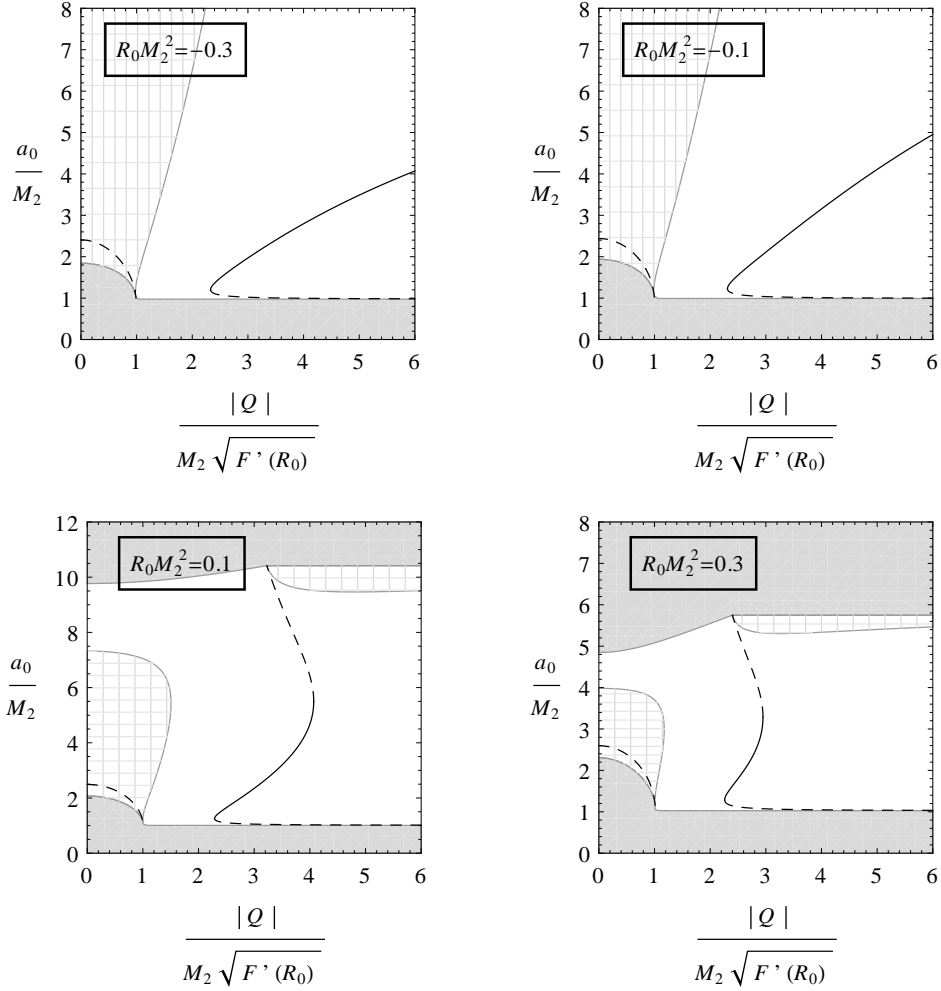}
\caption{ Shells around black holes in $ F (R) $ gravity, for different values of the curvature scalar $ R_0 $. The solid lines represent the stable static solutions with radius $ a_0 $, while the dashed lines the unstable ones. The masses satisfy the relation $ M_1 / M_2 = 0.5 $ and $ \alpha / M ^ 2_2 = 0.1 $ is taken. The charge $ Q $ corresponds to the exterior geometry. The meshed zones represent normal matter, while the gray areas have no physical meaning.}
\label{fig1}
\end{figure}

In order to start the construction of the shell we choose a radius $ a $ satisfying Eq. (\ref{CondGen}), larger than the event horizon radius in $ \mathcal{M}_1 $, so the black hole is always present, and when $ R_0> 0 $, also smaller than the cosmological horizon radius coming from the original geometry for this region. On the other hand, this radius $ a $ should be large enough to avoid the presence of the event horizon and the singularity of the geometry used for the region $ \mathcal{M}_2 $. When $ R_0> 0 $, it also has to be smaller than the cosmological horizon of this outer region. As we mentioned in the previous section, it is necessary that $ F '(R_0)> 0 $ to avoid ghosts. It is also preferable that the matter on the shell satisfies the weak energy condition, in order to guarantee the presence of normal matter on $ \Sigma $. The energy density and the pressure are given by Eqs. (\ref{e9}) and (\ref{e10}), respectively, which fulfill the equation of state (\ref{e11}).
 
In the static case, the radius $ a_0 $ should satisfy Eq. (\ref{CondEstatico}), while $ \sigma_0 $ and $ p_0 $ are given by Eqs. (\ref{e13}) and (\ref{e14}). The potential (\ref{potencial}) allows for the stability analysis of the solutions, which is determined from the sign of $ V '' (a_0) $, given by Eq. (\ref{potencial2der}); the stable ones correspond to $ V '' (a_0)> 0 $. The results are presented graphically in Fig. \ref{fig1}, in which the most representative ones are shown. All quantities are adimensionalized with the mass $ M_2 $ of the outer region; the relations $M_1 / M_2 = 0.5 $ and $ \alpha / M ^ 2_2 = 0.1 $ are adopted in all plots. The meshed zones represent those that satisfy the weak energy condition, and the gray areas have no physical meaning. Solid lines represent stable solutions, while unstable solutions are drawn with dotted lines. The behavior of the solutions does not vary significantly with the modulus of the value of the curvature scalar $ R_0 $, but mainly with its sign, obtaining that
\begin{itemize}
\item When $ R_0 <0 $, for $ |Q| <Q_ {c} $, there is only an unstable solution composed by normal matter. For larger values of $ |Q| $, there are two solutions made of exotic matter, one of them is stable while the other, close to the event horizon of the black hole, is unstable.
\item For $ R_0> 0 $ and $ | Q | <Q_ {c} $ there is an unstable solution constituted by normal matter. For values $ | Q |> Q_ {c} $ and a restricted range of charge, there are three solutions composed by exotic matter, one of them is stable. For values of $ |Q| $ much larger than $ Q_c $, there is only an unstable solution close to the event horizon of the black hole.
\end{itemize}
The function $ F (R_0) $, which is present through its derivative, does not produce significant changes in the qualitative behavior of the solutions, it only affects them by modifying their scale. The quotient $ | Q | / \sqrt{F '(R_0)} $ can be interpreted as an effective charge.

\subsection{Curvature scalars $R_1 \neq R_2$}

\begin{figure} [t!]
\centering
\includegraphics[width=0.8\textwidth]{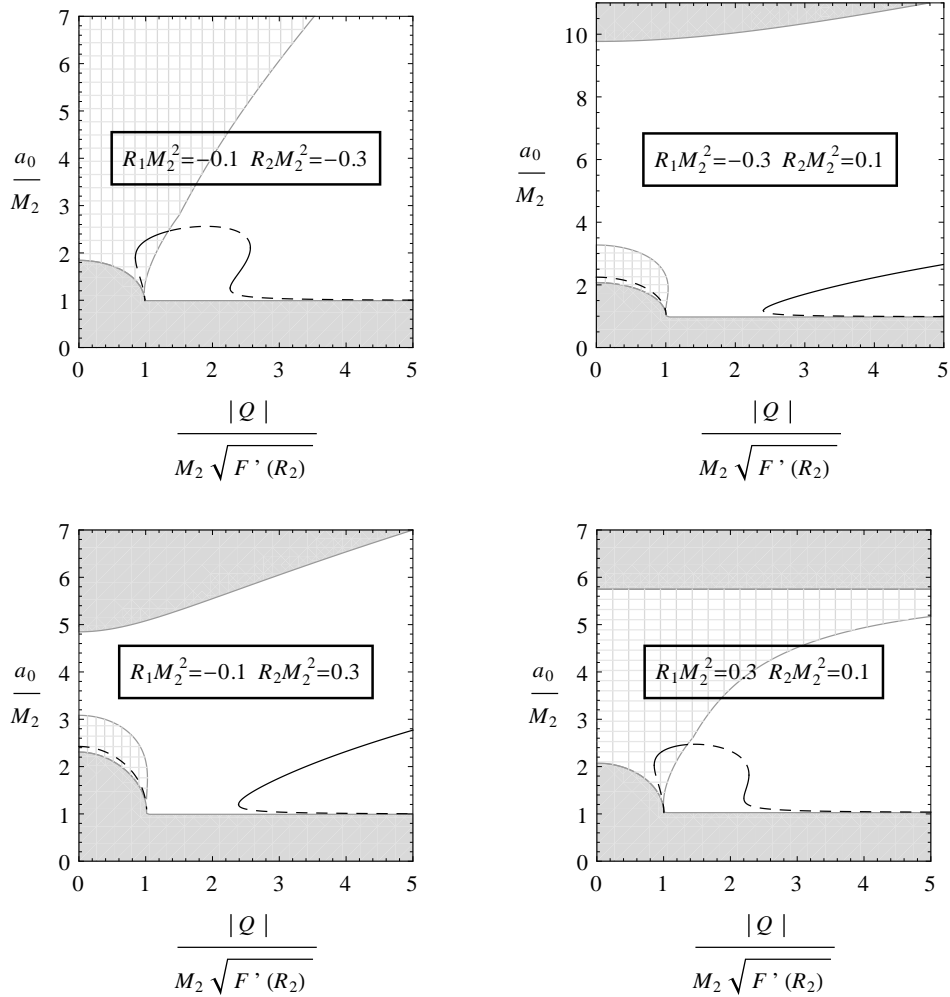}
\caption{ Shells around black holes in $ F (R) $ gravity, for different values of curvature scalars $ R_1 $ and $ R_2 $. The solid lines represent the stable static solutions with radius $ a_0 $, while the dashed lines the unstable ones. The mass ratio used is $ M_1 / M_2 = 0.5 $ and $ \alpha / M ^ 2_2 = 0.1 $ is taken. The charge $ Q $ corresponds to the exterior geometry. The meshed zones represent normal matter and the gray areas have no physical meaning.}
\label{fig2}
\end{figure}

\begin{figure}[t!]
\centering
\includegraphics[width=0.35\textwidth]{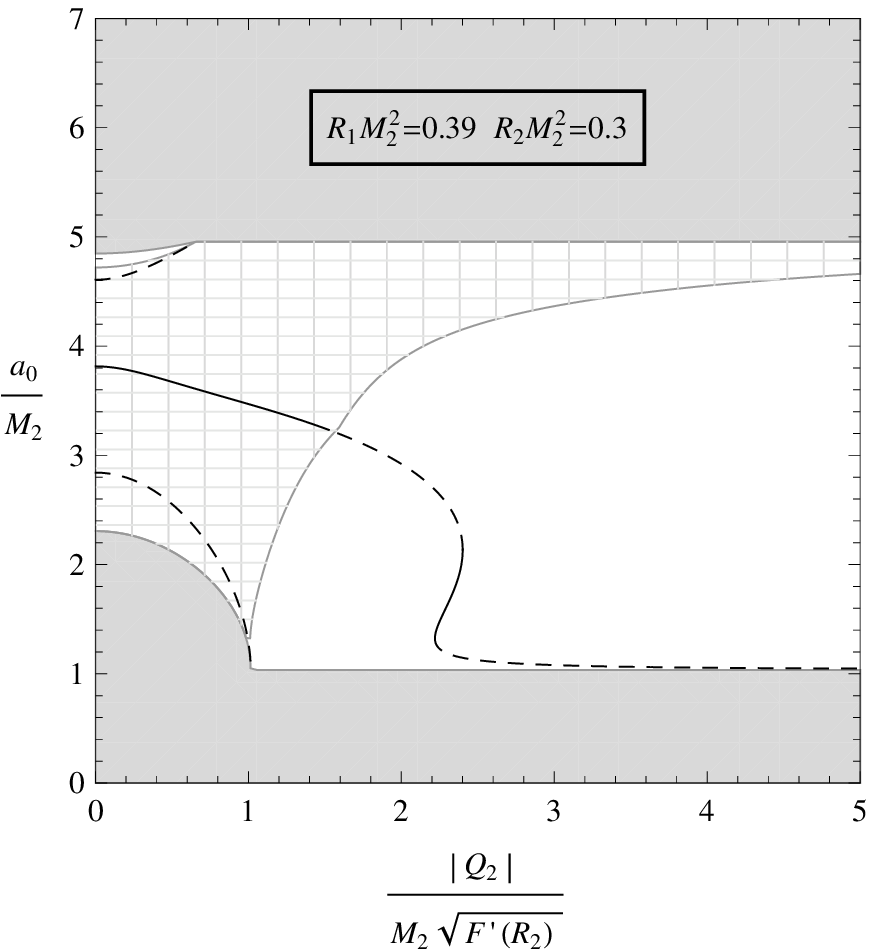}
\caption{ Idem Fig. \ref{fig2}, but with $ R_1 = 0.39 $ and $ R_2 = 0.3 $. We observe that there is a stable solution with normal matter and null charge at the shell.}
\label{fig3}
\end{figure} 

Analogously to the previous sub-section, we construct the shell by taking the mass $ M_1\neq 0 $ and a null charge for the internal region $ \mathcal{M}_1 $, and the mass $ M_2 \neq 0 $ and the charge $ Q $ for the external one $ \mathcal{M}_2 $. But now, we adopt different curvature scalars at the sides of the shell $ \Sigma $, so that $ [R]\neq 0 $. Then, the metric functions are
\begin{equation} 
A _1(r) = 1-\frac{2M_1}{r}-\frac{R_1 r^2}{12}
\label{A1_metric_Rdif}
\end{equation}
and
\begin{equation} 
A _2(r) = 1-\frac{2M_2}{r}+\frac{Q^2}{ F'(R_2) r^2}-\frac{R_2 r^2}{12}.
\label{A2_metric_Rdif}
\end{equation}
The possible horizons are found as explained in the previous sub-section, with the replacement of $R_0$ by $R_1$ or $R_2$ as appropriate, retaining the same characteristics described therein.

The radius $a$ of the shell is properly chosen, in the same way as done in the previous sub-section, and it should satisfy Eq. (\ref{CondGen}). The surface energy density is obtained by Eq. (\ref{enden}), the pressure by Eq. (\ref{pres}) and the external tension/pressure by Eq. (\ref{Trewrit}). These three equations determine, together with Eq. (\ref{CondGen}), the equation of state at the shell (\ref{state}). We also have that $ \mathcal{T}_\mu = 0 $ and, because $ [R]\neq 0 $, the non-null tensor $ \mathcal{T}_{\mu\nu} $, with a dipolar density $ \mathcal{P}_{\mu \nu} $ given by Eq. (\ref{dlsob}).

In the static case, the construction of the shell is done by choosing a radius $ a_0 $ that satisfies the Eq. (\ref{CondEstatico}). Besides, we assume that $F'(R_1)>0$ and $F'(R_2)>0$ in order to avoid the presence of ghosts. Again, we prefer matter fulfilling the weak energy condition. The surface energy density, pressure, and external tension/pressure are obtained from Eqs. (\ref{enden0}), (\ref{pres0}), and (\ref{T0}), respectively. As it is mentioned above, a non-null dipolar distribution given by Eq. (\ref{dlsob}) is also present. The stability of the solutions is determined by using the Eq. (\ref{potencial2der}), through the study of the sign of $ V '' (a_0) $, which guarantees the stability of the solution when $ V '' ( a_0)> 0 $. The results are displayed in Fig. \ref{fig2}. Again, the quantities are adimensionalized with the mass $ M_2 $, and the relations $ M_1 / M_2 = 0.5 $ and $ \alpha / M^2_2 = 0.1 $ are used. The meshed zones are those where the solution is composed by normal matter, while the gray ones have no physical meaning. The solid lines represent the stable solutions and the dotted ones the unstable ones. The behavior of the solutions basically depends on the relation between the values of the curvature scalars $ R_1 $ and $ R_2 $, the main features are
\begin{itemize}
\item For $ R_1 > R_2 $ and values of $ |Q| $ close to $ Q_c $, there are two solutions, one stable and the another unstable, both made of normal matter. For larger values of $ |Q| $, unstable solutions constituted by exotic matter predominate. Only for a restricted range of  $ |Q| $ there is a stable solution consisting of exotic matter.
\item For $ R_1 < R_2 $ and charge values $ | Q | < Q_c $, there is an unstable solution  made of normal matter. For $ | Q | > Q_c $ and for a broad range of values of charge, there are two solutions composed by exotic matter, one of them stable.
\end{itemize}
Again, the quotient $ | Q | / \sqrt{F '(R_2)} $ can be understood as an effective charge. In particular, with a suitable choice of parameters, i.e. $ R_1 = 0.39 $ and $ R_2 = 0.3 $, we have found that a stable solution made of normal matter and without charge is possible, as shown in Fig. \ref{fig3}.

\section{Conclusions}\label{conclu}

In this work, we have studied spherically symmetric thin shells of matter around black holes within $ F (R) $ theories of gravity. We have adopted constant curvature scalars at both sides of the shell and we have analyzed two scenarios: one in which the curvature scalars are equal to the same value $ R_0 $, and the other in which the values of the curvature scalars $ R_1 $ and $ R_2 $ are different. The case with both curvature scalars equal to $ R_0 $ does not impose any limitations on the function $ F (R) $. The matter at the shell should fulfill the equation of state $ \sigma - 2p = 0 $ that relates the surface energy density $\sigma $ and the pressure $p$. When the curvature scalars are different, it is necessary to restrict the analysis to quadratic $ F (R) $. Then, for $ R_1 \neq R_2 $ the shell is composed by matter that satisfies the equation of state $ \sigma - 2p = \mathcal{T} $, which depends on the external tension/pressure $\mathcal{T}$; it also requires the presence of the vector $ \mathcal{T}_\mu = 0 $ and the tensor $ \mathcal{T} _{\mu\nu} \neq 0 $ contributions. In particular, we have constructed a thin shell of matter surrounding a static non-charged black hole with mass $ M_1 $. The geometry outside the shell corresponds to a solution with mass $ M_2 $ and charge $ Q $.

In the case with the same value of the curvature scalar at both sides of the shell, the behavior of the solutions is determined by the sign of $ R_0 $, and it is always possible to find unstable solutions with normal matter. On the other hand, stable solutions constituted by normal matter are not found; those that are stable are present, but always built with exotic matter, within a small range of $ |Q| $ when $ R_0> 0 $ or for large values of $ |Q| $ when $ R_0 < 0 $. This result is similar to the one obtained in Ref. \cite{gfa18} for bubbles, with the main difference being that for the shells around black holes there is an extra solution, unstable and constituted by exotic matter, with values of the shell radius close to the event horizon one.

For different values of the curvature scalar at the regions separated by the shell, the solutions have a distinct behavior depending on the relation between $ R_1 $ and $ R_2 $. Only in the case with $ R_1> R_2 $ it is possible to find stable solutions constituted by normal matter for $|Q|$ close to $ Q_c $. The rest of the solutions are unstable and made of exotic matter, with the exception of a small range of $|Q|$ for which the shell is stable but, once again, it is composed by exotic matter. When $ R_1 <R_2 $, we have only found unstable solutions constituted by normal matter for $ |Q| < Q_c $, or by exotic matter for $|Q| > Q_c $. The only stable solution for this case is made of exotic matter for a broad range of values of the charge. These results are similar to those found in Ref. \cite{gfa18} in the case of $ R_1 \neq R_2 $, with the difference that, in shells surrounding black holes, an additional unstable solution emerges close to the event horizon. Compared with the case of the Ref. \cite{gfa18}, it is more complicated to find an appropriate set of parameters that allow the construction of a stable shell around the black hole, made of normal matter and without charge. However, we have found that it is possible to build such a case, for a limited range of values of $ R_1 $ and $ R_2 $, and we have shown an example with $ R_1 = 0.39 $ and $ R_2 = 0.3 $.

It is well known that there is an equivalence between any $F(R)$ gravity theory and a properly taken scalar--tensor theory \cite{dft,sofa}; in particular, quadratic $F(R)$ is equivalent to Brans-Dicke theory with a parameter $\omega =0$, with a potential $V(\phi)= 2 \Lambda+(\phi ^2 -2\phi -3)/(4\alpha)$, where the scalar field $\phi$ and the curvature scalar are related by $\phi=2\alpha R-1$. Then, it is worthy to note that the results obtained here can be translated to the corresponding scalar--tensor theory.

\section*{Acknowledgments}

This work has been supported by CONICET and Universidad de Buenos Aires.


\begin{thebibliography}{99}

\bibitem{sofa} T.P. Sotiriou and V. Faraoni, Rev. Mod. Phys.  \textbf{82}, 451 (2010).

\bibitem{dft} A. De Felice and S. Tsujikawa, Living Rev. Relativity \textbf{13}, 3 (2010).

\bibitem{nojod} S. Nojiri and S.D. Odintsov, Phys. Rep. \textbf{505}, 59 (2011); S. Nojiri, S.D. Odintsov, and  V.K. Oikonomou, Phys. Rep. \textbf{692}, 1 (2017).

\bibitem{bhfr1} T. Clifton and J.D. Barrow, Phys. Rev. D \textbf{72}, 103005 (2005); T. Multam\"aki and I. Vilja, Phys. Rev. D \textbf{74}, 064022 (2006); S. Capozziello, A. Stabile, and A. Troisi, Class. Quantum Gravity \textbf{25}, 085004 (2008).

\bibitem{bhfr2}  A. de la Cruz-Dombriz, A. Dobado, and A.L. Maroto, Phys. Rev. D  \textbf{80}, 124011 (2009); \textbf{83}, 029903(E) (2011).

\bibitem{moon} T. Moon, Y.S. Myung, and E.J. Son, Gen. Relativ. Gravit. \textbf{43}, 3079 (2011).

\bibitem{bhfr3} L. Sebastiani and S. Zerbini, Eur. Phys. J. C \textbf{71}, 1591 (2011); Z. Amirabi, M. Halilsoy, and S. Habib Mazharimousavi, Eur. Phys. J. C \textbf{76}, 338 (2016).

\bibitem{bhfr4} S. Nojiri and S.D. Odintsov, Class. Quantum Gravity \textbf{30}, 125003 (2013); S. Nojiri and S.D. Odintsov, Phys. Lett. B \textbf{735}, 376 (2014).

\bibitem{whfr} F.S.N. Lobo and M.A. Oliveira, Phys. Rev. D \textbf{80}, 104012 (2009);  A. DeBenedictis and D. Horvat, Gen. Relativ. Gravit. \textbf{44}, 2711 (2012); T. Harko, F.S.N. Lobo, M.K. Mak, and S.V. Sushkov, Phys. Rev. D \textbf{87}, 067504 (2013).

\bibitem{branefr} S. Chakraborty and S. SenGupta, Eur. Phys. J. C  \textbf{75}, 11 (2015).

\bibitem{daris} G. Darmois, M\'{e}morial des Sciences Math\'{e}matiques, Fascicule XXV, Chap. V (Gauthier-Villars, Paris, 1927); W. Israel, Nuovo Cimento B \textbf{44}, 1 (1966); \textbf{48}, 463(E) (1967).

\bibitem{sh1} P.R. Brady, J. Louko and E. Poisson, Phys. Rev. D \textbf{44}, 1891 (1991); M. Ishak and K. Lake, Phys. Rev. D \textbf{65}, 044011 (2002); S.M.C.V. Gon\c{c}alves, Phys. Rev. D \textbf{66}, 084021 (2002); F.S.N. Lobo and P. Crawford, Class. Quantum Gravity \textbf{22}, 4869 (2005).

\bibitem{sh2} E.F. Eiroa and C. Simeone, Phys. Rev. D \textbf{83}, 104009 (2011);  E.F. Eiroa and C. Simeone, Int. J. Mod. Phys. D \textbf{21}, 1250033 (2012); E.F. Eiroa and C. Simeone, Phys. Rev. D  \textbf{87}, 064041  (2013).

\bibitem{sh3} S.W. Kim, J. Korean Phys. Soc.  \textbf{61}, 1181 (2012); M. Sharif and S. Iftikhar, Astrophys. Space Sci., \textbf{356}, 89 (2015).

\bibitem{gravstar1} M. Visser and D.L. Wiltshire, Class. Quantum Gravity \textbf{21}, 1135 (2004);  N. Bili\'c, G.B. Tupper, and R. D. Viollier, J. Cosmol. Astropart. Phys.  \textbf{02}, 013 (2006).

\bibitem{gravstar2} F. S. N. Lobo and A. V. B. Arellano, Class. Quantum Gravity \textbf{24}, 1069 (2007); P. Martin-Moruno, N. Montelongo Garcia, F.S.N. Lobo, and M. Visser, J. Cosmol. Astropart. Phys.  \textbf{03}, 034 (2012).

\bibitem{wh1} E. Poisson and M. Visser, Phys. Rev. D \textbf{52}, 7318 (1995).

\bibitem{wh2}E.F. Eiroa and G.E. Romero, Gen. Relativ. Gravit. \textbf{36}, 651 (2004); F.S.N. Lobo and P. Crawford, Class. Quantum Gravity \textbf{21}, 391 (2004); G.A.S. Dias and J.P.S. Lemos, Phys. Rev. D \textbf{82}, 084023 (2010); V. Varela, Phys. Rev. D \textbf{92}, 044002 (2015).

\bibitem{wh3} E.F. Eiroa, Phys. Rev. D \textbf{78}, 024018 (2008); N. Montelongo Garcia, F.S.N. Lobo, and M. Visser, Phys. Rev. D \textbf{86}, 044026 (2012).

\bibitem{whcil}  E.F. Eiroa and C. Simeone, Phys. Rev. D \textbf{81}, 084022 (2010); \textbf{90}, 089906(E) (2014); S. Habib Mazharimousavi, M. Halilsoy, and Z. Amirabi, Phys. Rev. D  \textbf{89}, 084003 (2014); E.F. Eiroa and C. Simeone, Phys. Rev. D \textbf{91} 064005 (2015).

\bibitem{dss} N. Deruelle, M. Sasaki, and Y. Sendouda, Prog. Theor. Phys. \textbf{119}, 237 (2008).

\bibitem{js1} J.M.M. Senovilla, Phys. Rev. D \textbf{88}, 064015 (2013).

\bibitem{js2-3} J.M.M.  Senovilla, Class. Quantum Gravity \textbf{31}, 072002 (2014); J.M.M.  Senovilla, J. Phys. Conf. Ser. \textbf{600}, 012004 (2015).

\bibitem{js4} B. Reina, J.M.M. Senovilla, and R. Vera, Class. Quantum Gravity \textbf{33}, 105008 (2016).

\bibitem{eirfig}  E.F. Eiroa and G. Figueroa Aguirre, Eur. Phys. J. C \textbf{76}, 132 (2016);  E.F. Eiroa and G. Figueroa Aguirre, Phys. Rev. D \textbf{94}, 044016 (2016).

\bibitem{whfrts} M. Zaeem-ul-Haq Bhatti, A. Anwar, and S. Ashraf, Mod. Phys. Lett. A \textbf{32}, 1750111 (2017); S. Habib Mazharimousavi, Eur. Phys. J. C \textbf{78}, 612 (2018).

\bibitem{bb1} E.F. Eiroa, G. Figueroa Aguirre, and J.M.M. Senovilla, Phys. Rev. D \textbf{95}, 124021 (2017).

\bibitem{gfa18} E.F. Eiroa and G. Figueroa Aguirre, Eur. Phys. J. C \textbf{78}, 54 (2018).

\bibitem{bronnikov} K.A. Bronnikov, M.V. Skvortsova, and A.A. Starobinsky, Grav. Cosmol. \textbf{16}, 216 (2010).

\end{thebibliography}
\end{document}